\documentclass{moriond}

\def\be{\begin{equation}}
\def\ee{\end{equation}}
\def\bea{\begin{eqnarray}}
\def\eea{\end{eqnarray}}

\newcommand{\Photo}{}

\begin{document}
\vspace*{4cm}
\title{Recent measurements of strangeness and heavy flavor at STAR}

\author{A.G. Knospe (on behalf of the STAR Collaboration)}

\address{Department of Physics, Lehigh University, 16 Memorial Drive East,\\
Bethlehem, Pennsylvania 18015, USA}

\maketitle\abstracts{
The STAR Collaboration has collected collision data at a wide variety of center-of-mass energies and with several different species of colliding ions ($p$+$p$, Au+Au, Ru+Ru, Zr+Zr, and O+O). This data set enables a many studies of the properties of the hot and dense matter produced in ion-ion collisions, using a variety of probes. In these proceedings, recent STAR measurements of strangeness and heavy flavor production in ion-ion collisions are discussed.
}

\section{Strangeness Production}

Strangeness production is a sensitive probe of the properties of the medium created in heavy-ion collisions, providing insight into strangeness equilibration, hadronization dynamics, and the interplay between partonic and hadronic degrees of freedom. Measurements across different collision energies and system sizes allow the transition from hadronic to partonic dominated matter to be explored. STAR has measured the production of strange particles in Au+Au collisions at multiple center-of-mass energies as part of its fixed-target program. Strange hadron yields in Au+Au collisions at $\sqrt{s_{NN}}=3$~GeV were reported in~\cite{STAR_strangeness3GeV}. The production of singly strange hadrons and $\phi$ mesons exhibits a common scaling behavior (ref.~\cite{STAR_strangeness3GeV}, Fig. 6): the yields of $K^{-}$, $K^{0}_{S}$, and $\Lambda$ evolve $\propto\left\langle N_{\rm part}\right\rangle^{\alpha}$, with the power $\alpha$ being consistent among the three species. Here, $\left\langle N_{\rm part}\right\rangle$ is the mean number of nucleons that participate in collisions in the given centraltiy class. In contrast, the yields of protons and $\Xi^{-}$ scale differently with $\left\langle N_{\rm part}\right\rangle$. At this collision energy, protons mostly come from baryon stopping. The $\alpha$ value for $\Xi^{-}$ appears to be larger compared to $\Lambda$, $K_S^0$ and $K^{-}$, implying a stronger centrality dependence. This indicates that $\Xi$ production is likely driven by multi-step hadronic processes at $\sqrt{s_{\rm NN}}=3$~GeV, whereas this energy is below the threshold for direct $\Xi$ production in $NN$ collisions.

\begin{figure}
\centerline{\includegraphics[width=0.85\linewidth]{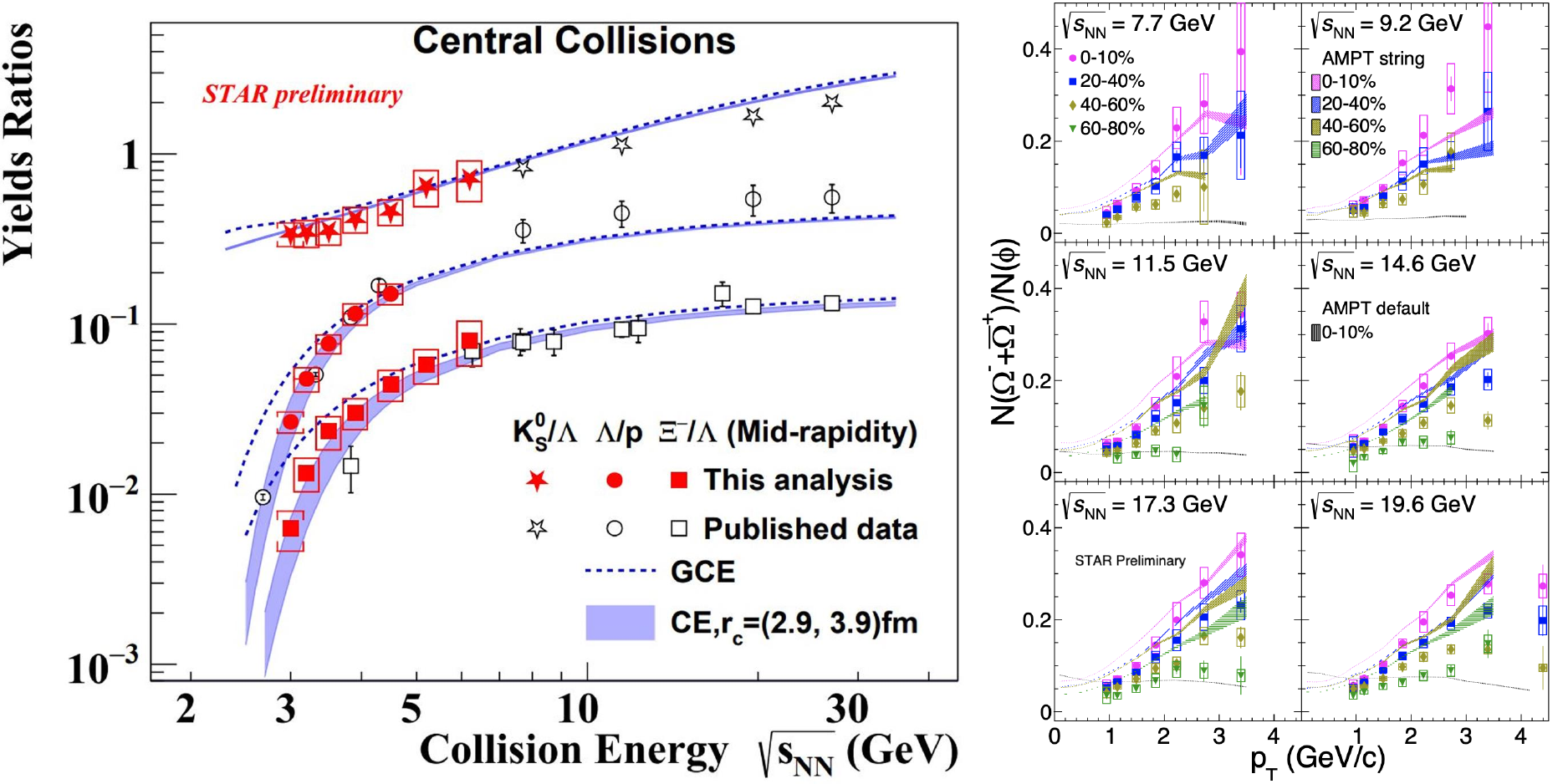}}
\caption[]{\textbf{Left:} Hadron yield ratios in central Au+Au collisions as a function of $\sqrt{s_{NN}}$~\cite{STAR_strangeness3GeV,Trzeciak_2025}, along with statistical model calculations~\cite{Vovchenko_2016,THERMUS}. \textbf{Right:} $\Omega/\phi$ yield ratios in Au+Au collisions as functions of $p_{\rm T}$ for different collision energies and centrality classes. Also shown are AMPT model calculations~\cite{AMPT}.}
\label{fig:strangeness}
\end{figure}

Figure~\ref{fig:strangeness}-left shows a study of strange hadron production in Au+Au collisions at multiple energies in the range $3\leq\sqrt{s_{NN}}\leq6.2$~GeV~\cite{Trzeciak_2025}. The $\Lambda/p$ and $\Xi^{-}/\Lambda$ yield ratios both exhibit sharp drops with decreasing $\sqrt{s_{NN}}$ at the lowest RHIC collision energies. This is due to (1) the presence of the elementary $NN$ production threshold (at $\sqrt{s_{NN}}\approx2.55$~GeV for $\Lambda$ and 3.25~GeV for $\Xi^{-}$), (2) the increase of proton yields with decreasing $\sqrt{s_{NN}}$ due to baryon stopping, and (3) canonical suppression of strange hadron production at low $\sqrt{s_{NN}}$. The Grand Canonical Ensemble (GCE) fails to describe the $\Lambda/p$ and $\Xi^{-}/\Lambda$ ratios for $\sqrt{s_{NN}}<5$~GeV, while the data are described by a Canonical Ensemble calculation with strangeness correlation radii in the range $2.9\leq r_{\rm C}\leq3.9$~fm~\cite{Vovchenko_2016,THERMUS}. This indicates that local strangeness conservation is important at the lowest RHIC collision energies. The increase in the $K^{0}_{S}/\Lambda$ ratio indicates a transition from baryon-dominated matter to meson-dominated matter as the collision energy increases.

The production of $\phi$ mesons and $\Omega$ baryons has been measured in Au+Au collisions at multiple energies in the range $7.7\leq\sqrt{s_{NN}}\leq19.6$~GeV; see Fig.~\ref{fig:strangeness}-right. The $\Omega/\phi$ yield ratio is observed to be enhanced for $p_{\rm T}>2$~GeV, with greater enhancement in more central collisions. This behavior, which has been observed at higher collision energies, can be attributed to radial flow and quark coalescence, providing evidence for partonic collectivity and hadronization via recombination mechanisms. A calculation using the default configuration of AMPT (without QGP) exhibit $p_{\rm T}$-independent $\Omega/\phi$ ratio, which is inconsistent with these STAR data~\cite{AMPT}. An AMPT calculation including string melting qualitatively reproduces the observed enhancement. These results are evidence for QGP formation in central Au+Au collisions for $\sqrt{s_{NN}}>7.7$~GeV.

The production of $K^{*0}$ mesons has been measured in Au+Au collisions at collision energies in the range $7.7\leq\sqrt{s_{NN}}\leq27$~GeV. The $K^{*0}/K$ yield ratio decreases with increasing collision centrality for all RHIC collision energies $>7.7$~GeV, and also at LHC energies (ref.~\cite{STAR_Kstar_BES}, Fig.~4). This suppression is attributed to the reduction of the measured $K^{*0}$ resonance yield due to rescattering of the $K^{*0}$ decay products in the hadron gas phase of the collision. The STAR and ALICE measurements of the $K^{*0}/K$ ratio follow a single trend as a function of $(dN_{\rm ch}/dy)^{1/3}$ (a proxy for the collision system's radius) for all collision energies $>62.4$~GeV. However, the lower measurements reported in~\cite{STAR_Kstar_BES} deviate below that trend: the suppression becomes stronger as $\sqrt{s_{NN}}$ decreases. This suggests differences in the types of hadronic interactions that are happening in the medium produced in these low-energy collisions, highlighting the sensitivity of short-lived resonances to the lifetime and dynamics of the hadronic phase. The suppression trends in both $\sqrt{s_{NN}}$ and collision centrality are well described by UrQMD calculations~\cite{Sahoo_UrQMD_2024}, although further studies are needed to clarify the nature of the suppression and its energy dependence.

\section{Heavy Flavor Production}

\begin{figure}
\includegraphics[width=\linewidth]{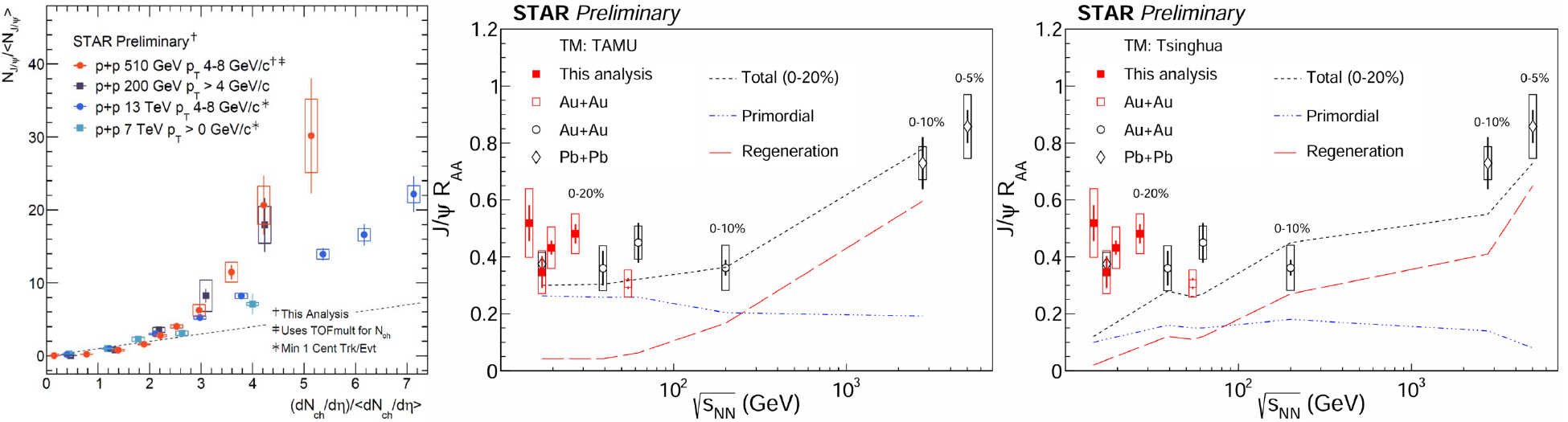}
\caption[]{\textbf{Left:} Self-normalized $J/\psi$ yield as a function of the mid-rapidity charged-particle multiplicity in $p$+$p$ collisions~\cite{Schaefer_psi}. \textbf{Center \& Right:} $J/\psi$ $R_{AA}$ in central Au+Au and Pb+Pb collisions as functions of $\sqrt{s_{NN}}$. Also shown are TAMU~\cite{TAMU_2010} (center) and Tsinghua~\cite{Tsinghua_2022} (right) model calculations.}
\label{fig:psi1}
\end{figure}

Studies of quarkonium production in $p$+$p$ collisions provide crucial baselines for comparison to $A$+$A$ collisions, and shed light on how quarkonia are produced. STAR has measured quarkonium production in $p$+$p$ collisions at $\sqrt{s}=510$~GeV (for $J/\psi$~\cite{Schaefer_psi}) and $\sqrt{s}=500$~GeV (for $\Upsilon(nS)$~\cite{STAR_Upsilon_pp500}) as a function of event activity. Faster-than-linear increases in the self-normalized $J/\psi$ (see Fig.~\ref{fig:psi1}-left) and $\Upsilon$ yields are observed as functions of the mid-rapidity charged-particle multiplicity; this behavior may be due to multi-parton interactions in the early stages of the collisions. For $J/\psi$, the trends measured at $\sqrt{s}=200$ and 510~GeV are consistent, but there are indications that those trends differ from measurements by ALICE at $\sqrt{s}=7$ and 13~TeV, with the ALICE data exhibiting a more gradual increase in $J/\psi$ yields with multiplicity. The $\Upsilon$ self-normalized yields follow similar trends at both RHIC and LHC energies (ref.~\cite{STAR_Upsilon_pp500}, Fig.~10); the increase is at least qualitatively described by PYTHIA8, the CGC/Saturation model, and the String-Percolation model (citations in ref.~~\cite{STAR_Upsilon_pp500}). STAR observes no statistically significant changes in the $\Upsilon(nS)/\Upsilon(1S)$ yield ratios as a function of event activity (ref.~\cite{STAR_Upsilon_pp500}, Fig.~8), suggesting that comover interactions do not have large effects on $\Upsilon$ yields in $p$+$p$ collisions.

Due to the larger spatial extent of the $\psi(2S)$ in comparison to the $J/\psi$, as well as its higher mass, the $\psi(2S)$ serves as a unique probe of the matter produced ion-ion collisions. The sequential suppression of charmonium has also been observed by STAR~\cite{STAR_psi2S}. The double yield ratio $\psi(2S)/(J/\psi)$ (the ratio in $A$+$A$ collisions, divided by the same ratio in $p$+$p$ collisions) has been measured for a combined Ru+Ru and Zr+Zr data set at $\sqrt{s_{NN}}=200$~GeV. It is observed that the excited $\psi(2S)$ is more suppressed than the ground state $J/\psi$ in $A$+$A$ collisions than in the $p$+$p$ baseline. The new STAR measurements are consistent with results from other accelerator facilities at both lower and higher collision energies. They also suggest stronger $\psi(2S)$ suppression in more central collisions. The suppression trend as a function of $\left\langle N_{\rm part}\right\rangle$ is reproduced by Tsinghua model and SHMc calculations (citations in ref.~~\cite{STAR_psi2S}). The $\psi(2S)/(J/\psi)$ double yield ratio is 3 standard deviations below the value for $p$+Au collisions, obtained through interpolation. This suggests that the observed $\psi(2S)$ suppression is due to the presence of hot nuclear matter.

STAR has conducted multiple measurements of quarkonium suppression as a function of collision energy and system size. This allows for studies of QGP properties under a wide variety of conditions, and sheds light on the relative importance dissociation, regeneration, and cold nuclear matter effects in determining quarkonium yields. STAR has recently released a measurement of the $J/\psi$ nuclear modification factor in Au+Au collisions at $\sqrt{s_{NN}}=54.4$~GeV~\cite{STAR_psi_54GeV}. The precision of this study is improved with respect to earlier measurements at $\sqrt{s_{NN}}=39$ and 62.4~GeV. For a given value of $\left\langle N_{\rm part}\right\rangle$, the measured $J/\psi$ $R_{AA}$ values across the RHIC energy range are consistent within uncertainties (ref.~\cite{STAR_psi_54GeV}, Fig.~3). In contrast, the $J/\psi$ $R_{AA}$ is significantly larger at LHC energies than at RHIC energies for $\left\langle N_{\rm part}\right\rangle>150$ due to charmonium regeneration. STAR's $J/\psi$ $R_{AA}$ measurements suggest stronger $J/\psi$ suppression in central Au+Au collisions and at low $p_{\rm T}$. The Tsinghua model~\cite{Tsinghua_2022} describes the observed suppression trend well as a function of $\left\langle N_{\rm part}\right\rangle$, with little dependence on $\sqrt{s_{NN}}$.

As shown in Fig.~\ref{fig:psi1}-right, STAR has measured the $J/\psi$ $R_{AA}$ at several collision energies (reaching as low as $\sqrt{s_{NN}}=17.3$~GeV) as part of its Beam Energy Scan program. These new low-energy $J/\psi$ $R_{AA}$ measurements are consistent with the value measured in Pb+Pb collisions at similar $\sqrt{s_{NN}}$. The $J/\psi$ $R_{AA}$ exhibits no significant dependence on $\sqrt{s_{NN}}$ across the RHIC energy range, while the larger regeneration effect is clearly visible at LHC energies. In the two panels, the experimental data are compared with TAMU~\cite{TAMU_2010} and Tsinghua~\cite{Tsinghua_2022} model calculations. In those calculations, the regeneration contribution (red long-dashed lines) is separated from the signal due to surviving primordial $J/\psi$ (blue dash-dotted lines). For both models, regeneration becomes the dominant source of $J/\psi$ for $\sqrt{s_{NN}}>250$~GeV. However, the two models differ regarding the relative sizes of the regeneration and primordial contributions at low $\sqrt{s_{NN}}$, and the observed trend is better described by the TAMU model. The Tsinghua model predicts $J/\psi$ $R_{AA}<0.2$ for $\sqrt{s_{NN}}<20$~GeV, which is inconsistent with the new STAR measurements.

\begin{figure}
\centerline{\includegraphics[width=0.75\linewidth]{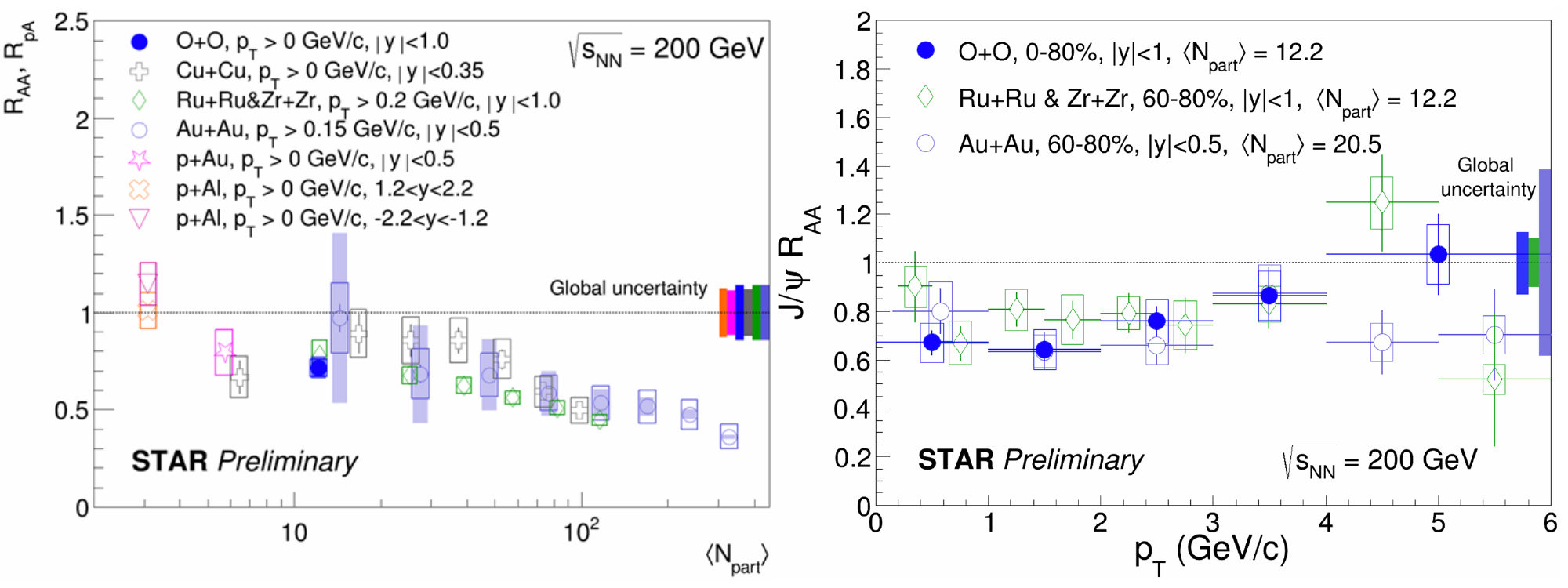}}
\caption[]{$J/\psi$ $R_{AA}$ in ion-ion collisions at $\sqrt{s_{NN}}=200$~GeV, plotted as functions of $\left\langle N_{\rm part}\right\rangle$ (left) and $p_{\rm T}$ (right).}
\label{fig:psiOO}
\end{figure}

Figure~\ref{fig:psiOO} shows new STAR measurements of $J/\psi$ $R_{AA}$ in O+O collisions at $\sqrt{s_{NN}}=200$~GeV, exhibiting strong suppression of $J/\psi$. The O+O data point lies along the trend in $\left\langle N_{\rm part}\right\rangle$ seen in other collision systems at the same energy. Production of $J/\psi$ is less suppressed in small collision systems than in large systems, with a smooth decrease in $J/\psi$ $R_{AA}$ observed as the system size increases. For O+O collisions, the STAR $J/\psi$ $R_{AA}$ increases as a function of $p_{\rm T}$, an effect not observed in peripheral Ru+Ru and Zr+Zr collisions (combined) and Au+Au collisions, with current uncertainties.

\section{Conclusion \& Acknowledgement}

The versatility of the RHIC complex enabled collisions of many different ion species and allowed for center-of-mass energies ranging from 3 to 200~GeV, with an upper limit of 510~GeV for $p$+$p$ collisions. This rich data set has allowed for many studies that probe the properties of strongly interacting matter as a function of collision energy and system size. Between 2023 and the end of RHIC running in early 2026, STAR recorded over $9\!\times\!10^{9}$ minimum-bias-triggered Au+Au collisions at $\sqrt{s_{NN}}=200$~GeV, plus over $4\!\times\!10^{9}$ high-luminosity or high-$p_{\rm T}$ triggered events. This data set will allow for many more studies in the coming years, using a variety of observables, with improved acceptance, uncertainties, kinematic reach, and overlap with the LHC.

The author acknowledges support from the U.S. Dept. of Energy, Grant \# DE-SC0023491.

\section*{References}

\begin{thebibliography}{99}
\bibitem{STAR_strangeness3GeV} M.I. Abdulhamid \textit{et al.} (STAR Collaboration), \textit{J. High Energy Phys.} \textbf{10} 139 (2024)
\bibitem{Trzeciak_2025} B. Trzeciak (STAR Collaboration), presentation at 25th Zim\'{a}nyi School (2025)
\bibitem{Vovchenko_2016} V. Vovchenko, V.V. Begun, and M.I. Gorenstein, \textit{Phys. Rev. C} \textbf{93} 064906 (2016)
\bibitem{THERMUS} S. Wheaton, J. Cleymans, M. Hauer, \textit{Comput. Phys. Commun.} \textbf{180} 84--106 (2009)
\bibitem{AMPT} T. Shao \textit{et al.}, \textit{Phys. Rev. C} \textbf{102} 014906 (2020)
\bibitem{STAR_Kstar_BES} STAR Collaboration, arXiv:2601.14884 (2026)
\bibitem{Sahoo_UrQMD_2024} A.K. Sahoo, S. Singha, and M. Nasim, \textit{J. Phys. G} \textbf{52} 015101 (2024)
\bibitem{Schaefer_psi} B. Schaefer (STAR Collaboration), \textit{EPJ Web of Conferences} \textbf{339} 04009 (2025)
\bibitem{STAR_Upsilon_pp500} B.E. Aboona \textit{et al.} (STAR Collaboration), \textit{Phys. Rev. D} \textbf{112} 032004 (2025)
\bibitem{STAR_psi2S} B.E. Aboona \textit{et al.} (STAR Collaboration), \textit{Phys. Rev. Lett.} \textbf{136} 122302 (2026)
\bibitem{STAR_psi_54GeV} STAR Collaboration, arXiv:2506.20962 (2025)
\bibitem{Tsinghua_2022} J. Zhao and P. Zhuang, \textit{Phys. Rev. C} \textbf{105} 064907 (2022)
\bibitem{TAMU_2010} X. Zhao and R. Rapp, \textit{Phys. Rev. C} \textbf{82} 064905 (2010)
\end{thebibliography}


\end{document}